\documentclass[pre,twocolumn,showpacs,preprintnumbers] {revtex4}

\usepackage{graphicx}
\usepackage{dcolumn}
\usepackage{bm}
\begin{document}
                                                                                                                             
\preprint{CUPhysics/09/2007}
\title
{Effect of a static phase transition on searching dynamics}

\author{Kamalika Basu Hajra and Parongama Sen}
\affiliation{
Department of Physics, University of Calcutta,92 Acharya Prafulla Chandra Road,
Calcutta 700009, India.\\
}

\begin{abstract}

We consider  a one dimensional Euclidean network which is grown using a preferential attachment. Here
the  $j$th incoming node gets attached to the $i$th existing node with the probability
$\Pi_i \propto k_i {{l}}_{ij}^\alpha$, where ${l}_{ij}$ is the
Euclidean distance between them and $k_i$ the degree of the $i$th node.
This network is known to have a static phase transition point at $\alpha_c \simeq  0.5$.
On this network, we employ three different searching strategies
based on  degrees or  distances or both, where the possibility of
termination of search chains is allowed. A detailed analysis
shows that these strategies are significantly affected by the presence
of the static critical point.
The distributions of the search path lengths and the success rates are also
estimated and compared for the different strategies. These distributions
appear to be marginally affected by the static phase transition.

\end{abstract}

\maketitle

\section{Introduction}

Static critical points are known to affect dynamical phemomena usually
giving rise to critical slowing down, e.g., as in the 
  relaxation of the order parameter
in magnetic systems. 
Dynamical critical phenomena is a well studied and important topic, exploring
the dynamical  behaviour of systems, especially at the thermal critical point \cite{Hohalp}.
In many  systems, phase transitions  
driven by factors other than temperature can occur, as for example 
the geometrical phase transition
occurring in percolation. Even in magnetic systems, e.g.,  in the  
axial next nearest
neighbour Ising (ANNNI) model, a phase transition occurs at zero termperature 
when the
competing second neighbour interaction takes up a certain value. However,
here a zero temperature  quenching dynamics fails to carry any 
signature of the phase transition \cite{redneretc}.

Recently, with the discovery of the small world effect 
in real  networks, many theoretical models have been set up
to mimic  small worlds. 
In some of these models, interesting phase transitions 
have been noted, e.g., in the Watts-Strogatz model \cite{watts,bareview}, where one
starts with nearest neighbour links only and then rewires links to long range neighbours with probability $p$, the small world
effect is observed even as $p\to 0$. Phase transitions
in models in which the linking probability is dependent on 
 spatial and/or  temporal factors have also been observed \cite{psmanna,kamalika1}. 
There is no temperature associated with these 
networks. Of course, if one considers spin systems, e.g., the Ising model,
on such networks, it is possible to obtain 
thermal phase transitions as well.  Dynamical studies of such systems 
at both zero and non-zero temperature   have  
shown unexpected phenomena, as for example freezing in 
case of  the quenching dynamics of the Ising model on small world networks \cite{freeze}.

The idea of a small world first emerged from a real dynamical
experiment made on the US population by Milgram \cite{milgram} which showed that
on an average there are six steps required to reach another individual.
Later, a mathematical definition of small world property was
proposed; by the   small world property it is  meant  that if
any two nodes in the network is separated by an average number of $s$ 
steps,
 then $s \propto \ln(N)$, where $N$ is the total number of 
nodes in the network.
In some networks, even slower variation (i.e., sub-logarithmic) scaling has been observed \cite{newman_sub}.

Following Milgram's original experiment, several  new experiments
have been done to
study the searching dynamics in real social networks \cite{killworth,dodds}.
A considerable number of theoretical studies on searching phenomena has
also been made recently
 \cite{adamic_search,hong,geog,
klein,adamic1,kim,zhu,moura,watts-search,carmi,thada,clauset,psen}.

While the small world property is a ``static'' property, i.e.,
calculated on the basis of global knowledge and without any dynamics 
involved, 
it must be noted that  it is not necessary that
a navigation or search on a small world 
network would  show the small world property, i.e., the dynamic 
paths  $s_d$ may not scale as  $\ln(N)$.
This is because searching is done using local 
information only.
This was explicitly shown by Kleinberg \cite{klein} in a theoretical study
where nodes where placed on a two dimensional Euclidean space. 
Each node here has connections to its  nearest neighbours  as well as to
neighbours at a distance $l$ with probability  
\begin{equation}
\label{ldist}
 P(l) \propto l^{-\alpha}.
\end{equation}
Although the network is globally a small world for a range of values of 
$\alpha$, 
navigation on such networks using greedy algorithm showed a
small world behaviour with $s_d \sim (log(N))^2$ only at $\alpha = 2$. 
In general the path length $s_d$ showed a sublinear power law increase with $N$.

Although  searching dynamics is not comparable to dynamics of relaxation, quenching etc.,
still we have some indication that it bears the signature of a static 
phase transition from  some earlier studies
  \cite{klein,zhu}. 
In one dimension, the Euclidean network in which connections
are made using (\ref{ldist})  has been shown 
to have three phases  corresponding to the scaling behaviour of the link 
lengths.
 It is a small world for 
for $\alpha < 1$  where
  the link lengths  scale as $N$; finite dimensional network for  $1< \alpha < 2$ where the
link lengths scale sublinearly with $N$ and a regular network for $\alpha>2$ where
the link lengths are $O(1)$ \cite{psarnab}. Here the existence of the two
phase transitions are quite confirmed and the greedy searching strategy 
has been shown to bear the signatures of  both \cite{zhu}.  
 Correspondingly, it is expected that in 
two dimensions, there is a transition to the small world behaviour at $\alpha_c=2$ and the
 result that the search path lengths scale uniquely at this value of $\alpha$ shows that the 
searching dynamics is indeed sensitive to it.
However, this has been  observed  for a particular algorithm and may not be 
true always. 
In this paper, we have considered a network where
different kind of algorithms can be used to check the sensitivity 
to  a static phase transition.

We have considered realistic searches on this network. As  in \cite{psen}, 
here also the
search paths may terminate. Thus there is a success rate also 
involved in the study. To study the effect of the parameter $\alpha$  which governs the 
phase transition in the model (details given in the next section), 
we have computed the path length, the success
rate and the ratio of the two as well. 
The last quantity  shows a power law variation with
the network size, $N^{-\delta}$,   and can be taken as a reliable measure
to compare different search strategies \cite{psen}.
We find out that the variation  of $\delta$ with $\alpha$
 indeed shows the signature of the phase transition to a considerable extent
 for the different algorithms.

We describe the model and the algorithms  in section II and the results in section III.
We have computed the distribution of the path lengths and success rates 
for all the algorithms  also which are presented in sec IV.
Summary and concluding remarks are made in sec V.

\section{The network and the strategies}

We have considered a growing Euclidean network in which the nodes are added one by one
using a preferential attachment such that the probability that an 
incoming node $j$ gets attached to an 
existing node $i$ with $k_i$ 
neighbours at that time is \cite{psmanna}
\begin{equation}
\label{attach}
\Pi_i \propto k {l}_{ij}^{-\alpha},
\end{equation}
where $l_{ij}$ is the distance between the $i$th and the $j$th nodes. 
The nodes are generated randomly at sites $x_i ~~(0 \leq x_i \leq 1$) 
on a continuous one dimensional line 
and each 
new node gets attached to $m$  pre-existing nodes.

For $m = 1$, this network was shown to have a phase transition at $\alpha_c$ : below $\alpha_c$ it has a stretched exponential
degree distrtibution while above it the degree distribution has a 
power law tail with the exponent $\gamma$ equal to  $3.0$. The value of $\alpha_c$ was found to be close to  $0.5$ \cite{psmanna}.

It is to be noted that this model, in the limit $\alpha =0$ is nothing
but the scale-free Barabasi Albert model \cite{BA}. On scale-free networks,
a degree based search is natural to adopt while  the
greedy algorithm appears to be the most popular one from the findings of 
the  original Milgram experiment as well as that of 
 \cite{killworth,dodds} on a Euclidean network. 
We have therefore employed three different search algorithms here, 
based on  the degrees of the nodes, or distances, or both.

All the  search strategies follow the basic rules :\\
After a  source node and a target  node are selected randomly, 
the source node will 
send the signal to one of its neighbouring nodes provided that node has not
already taken part in the search. This is repeated till the message reaches a node
which is connected to the target node and this scheme  is in tune with Milgram-like experiments. 
In course of this search, it may happen that a node cannot pass the signal to any of its
neighbour as they have already taken part in the search. In that case,
the search terminates at that node. Such searches are termed {\it unsuccessful}.
The fraction of successful searches  
is called the success rate $\rho$. The average number of 
steps taken to reach the target in a successful search is  the average dynamic path 
length $s_d$. We also calculate the quantity $\mu$, defined as $\mu = \frac{\rho}{s_d}$.

The choice of the neighbour to whom the signal is being passed 
depends on the strategy. The three search strategies considered  in the present 
work are as follows:\\

(1) Highest Degree Search (HDS): Here after a  source and a target
 pair are chosen randomly the source scans its nearest neighbours
 and chooses the one
with the highest degree to pass on the signal.

(2) Nearest neighbour search (NNS):  In this strategy, after the source-target pair is
chosen randomly, the source chooses from among its nearest neighbours, the
one whose Euclidean distance ($l$) from the target is the least.  
It may be noted that in conventional greedy algorithms, the strategy is to 
pass the message to a neighbour which is ${\it{nearer}}$ to the target than itself. 
In the present case, this condition has not been imposed and therefore
in an intermediate stage, the message may ``proceed backward''. This is analogous to
allowing a system to go a higher energy configuration in simulated annealing 
applied to the dynamics of
frustrated systems like spin glasses and
to combinatorial optimisation problems.

(3) Optimised Search (OS): In this strategy, we follow an algorithm where
the degree ($k$) of a node as well as its Euclidean distance ($l$) from the
target are taken into account. Here, after a source-target pair is picked up
at random, the ratio $\xi = k/l$ is calculated for all the nearest neighbours
 of the source and the one with the highest value of $\xi$ is
chosen to pass on the signal. 

\section{Results for $\rho, s_d$ and $\delta$}

We have simulated the networks with a maximum of  $N = 5000$ nodes using upto
 $1000$ configurations. For each configuration, the searching is repeated 
$N/2$ times with randomly chosen source-target pairs. 
We have considered two cases,  $m = 1$ and $m = 2$.
\subsection{Case I, m = 1}
Here the tunable  parameter $\alpha$ 
has been varied from $-10.0$ to $+10.0$.
 Once the network is generated following 
eq. (\ref{attach}), and the navigation has been simulated following one of the three 
strategies described in the last section, the success rate $\rho$ and the
 average search length $s_{d}$
 are evaluated and their variation with $\alpha$ and $N$  is noted.

\begin{center}
\begin{figure}
\noindent \includegraphics[clip,width= 5cm, angle=270]{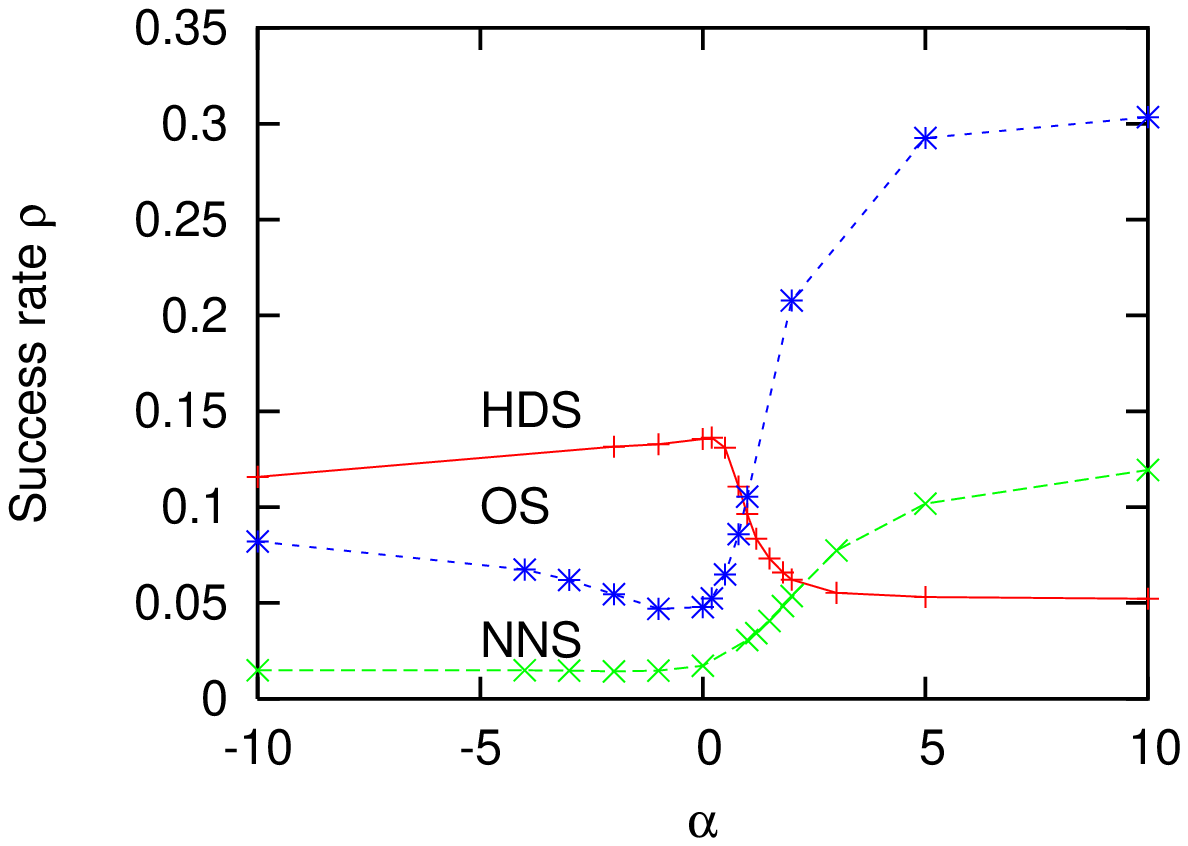}
\noindent \includegraphics[clip,width= 5cm, angle=270]{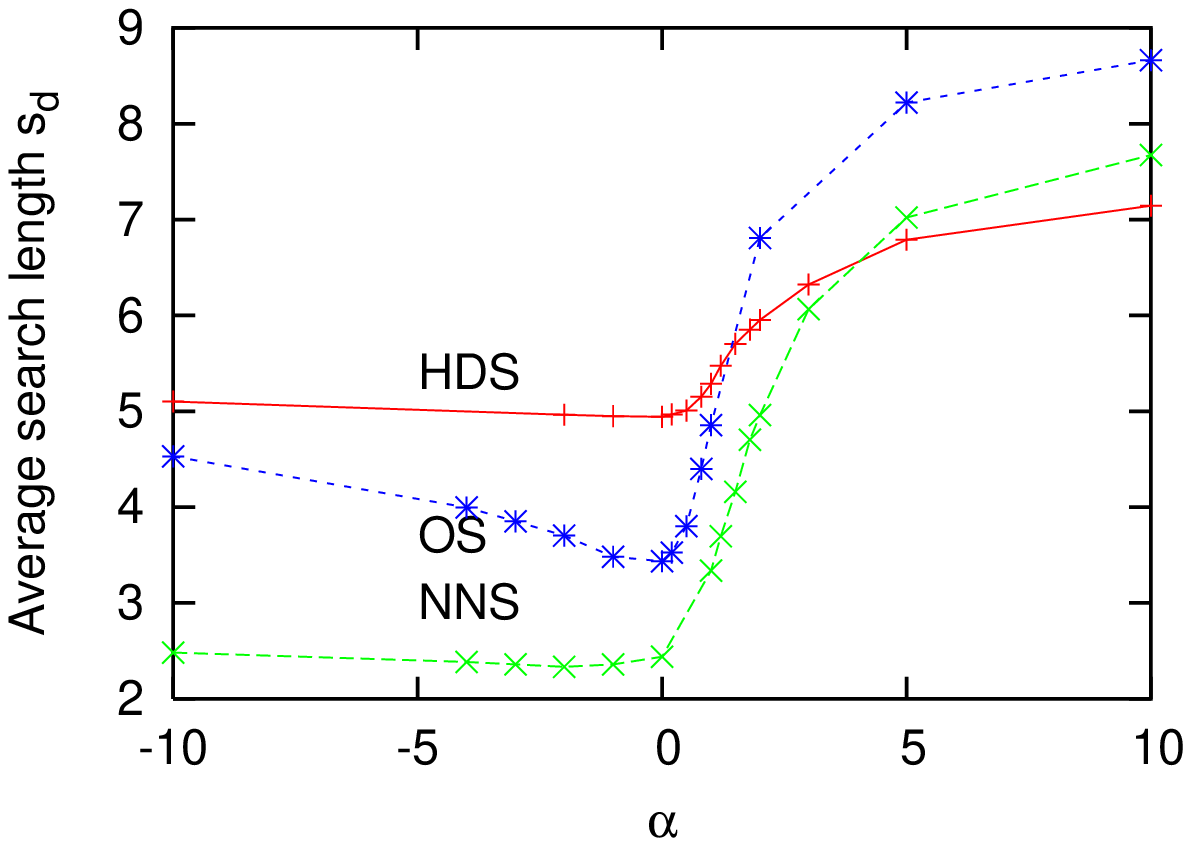}
\noindent \includegraphics[clip,width= 5cm, angle=270]{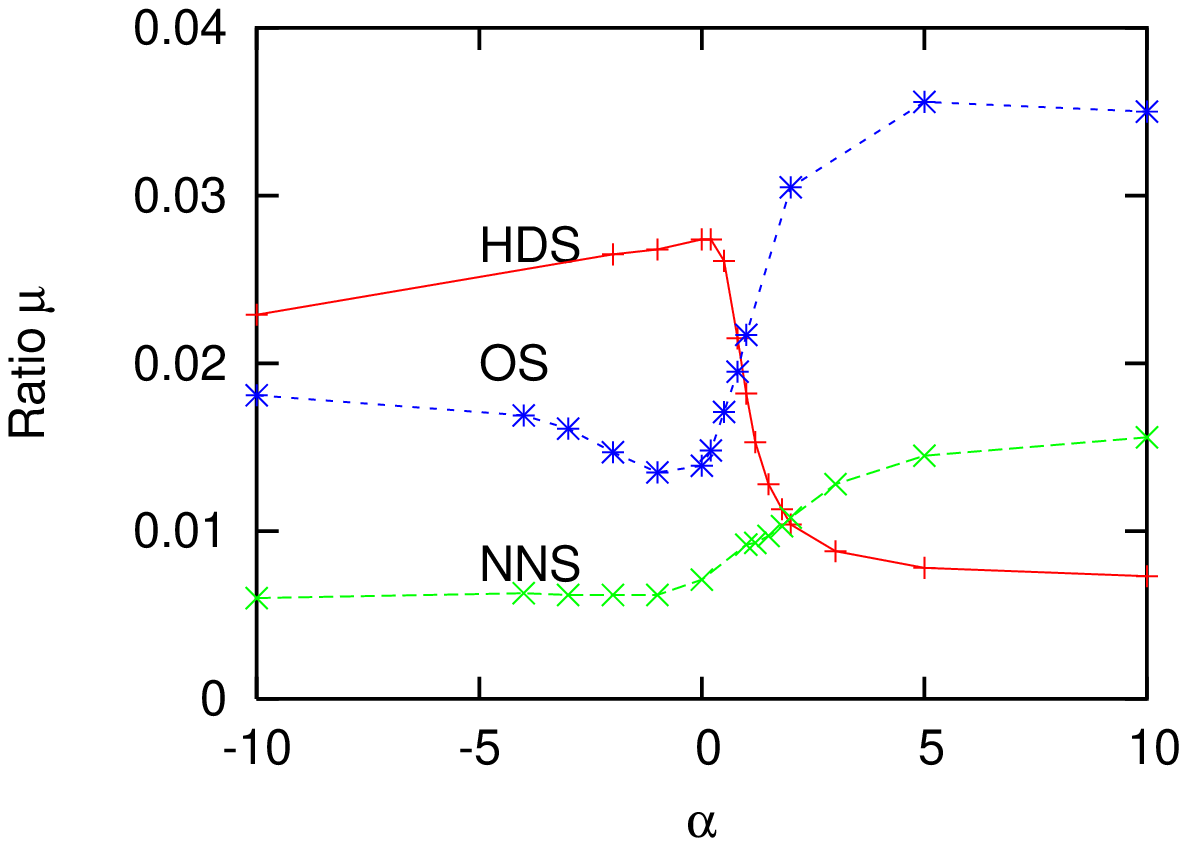}
\caption{Variation of $\rho$, $s_{d}$ and $\mu$ with $\alpha$ for the
 three search strategies  for $N = 1000$. The  parameter $\alpha$ is 
varied from $-10.0$ to $+10.0$. A transitional behaviour is observed for all 
the strategies around the static critical point of the system, i.e., 
near $\alpha = \alpha_c \simeq 0.5$}.
\end{figure} 
\end{center}

First we have made a  comparison of the three strategies
by analysing the variation 
of  $\rho$, $s_{d}$ and $\mu$ with  $\alpha$
 for a fixed system size $N$. We show in Fig. 1
 these variations for $N = 1000$.\\

For the HDS strategy, it is observed that $\rho$ shows a slow 
increase as $\alpha$ increases from $-10.0$ upto $0.0$ after which it falls 
sharply till $\alpha = 2.0$ and finally tends to saturate beyond $\alpha = 5.0$. The 
value of $s_{d}$ for this strategy however remains constant from $\alpha = -10.0$ to $\alpha = 0.0$ and increases slowly from this value  
also showing a tendency to saturate at large values of 
 $\alpha$. 

For the NNS strategy, $\rho$ and $s_d$ remain very small for $\alpha < 0$; $\rho$  
shows a gradual increase between 
$\alpha = 0.0$ and $\alpha = 2.0$. The values of $s_d$ however increase quite 
rapidly between  $\alpha = 0.0$ and $\alpha = 2.0$.  

For the OS strategy,  $\rho$ has a slow decrease from $\alpha = -10.0$ 
upto $\alpha = 0.0$ and then it  increases quite sharply between 
$\alpha = 0.0$ and 
$\alpha = 5.0$ beyond which it saturates.
Similarly  $s_{d}$ 
decreases slowly between $\alpha = -10.0$ and $\alpha = 0.0$, then increases
 very sharply upto $\alpha = 2.0$.

\begin{figure}
\noindent \includegraphics[clip,width= 5cm, angle=270]{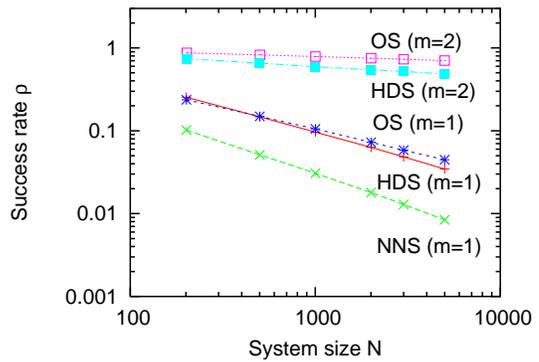}
\caption{Variations of $\rho$ with system size $N$ are shown for $\alpha = 1.0$ for 
the three search strategies for $m=1$ and $m=2$. The success rate decreases with increasing system
 size.}
\end{figure}
\begin{figure}
\includegraphics[clip,width= 5cm, angle=270]{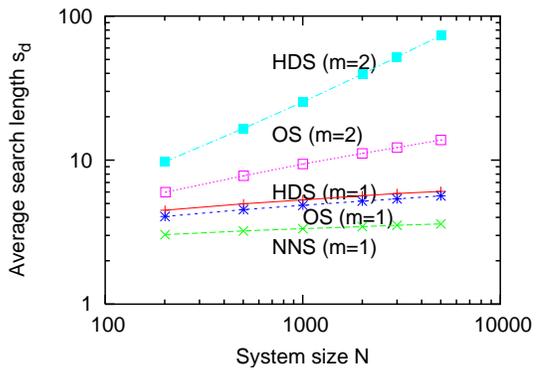}
\caption{Variations of the average path length $s_{d}$ with system size $N$ 
for the three different strategies are shown for $\alpha = 1.0$ for $m=1$ and 
$m=2$. It is observed that 
the average search length increases with increasing $N$.}
\end{figure}

A saturation of both $\rho$ and $s_d$  is expected for all the algorithms
 as the network approaches the behaviour  of a  
growing network in which links are made to the nearest neighbours
 (i.e., the $\alpha \to \infty$ limit) for large $\alpha$. Similar saturation 
behaviour for the static properties was observed in \cite{psmanna}\\

Looking at $s_{d}$ alone, it would seem that the HDS is still the best 
strategy even at $\alpha >> 1$, when the network
is not scale-free. However, $\rho$ for HDS becomes very low here indicating that  very few chains are completed,
 in which case  chains tend to be `short'.  
 This explains the above observation for $s_d$. 
Similarly, for $\alpha < 0$, NNS would seem to be 
the best from the values of $s_{d}$.
 On the other hand, from the $\rho$ plots, OS seems 
best for $\alpha >> 1$, while HDS seems best for $\alpha <0$. As in 
\cite{psen}, here also we compute $\mu$, which incorporates both
$\rho$ and $s_d$,  to comment on the relative
 capabilities of the three strategies. From the $\mu$ vs $\alpha$ plots above 
it is apparent that HDS works best upto $\alpha \sim 0.5$ while OS is best 
for $\alpha > 0.5$ for this particular value of $N$. NNS works rather poorly for
$\alpha <0$ and performs relatively better for $\alpha >0$.

We find that the behaviour of $\mu$ in general closely follows that of the 
success rates $\rho$. This may indicate that rather than the path lengths,
which are ``small'' in all cases, the success rate decides the quality of the
search strategy here.

Next we discuss the behaviour of the above quantities with $N$.

\begin{figure}
 \includegraphics[clip,width= 4cm, angle=270]{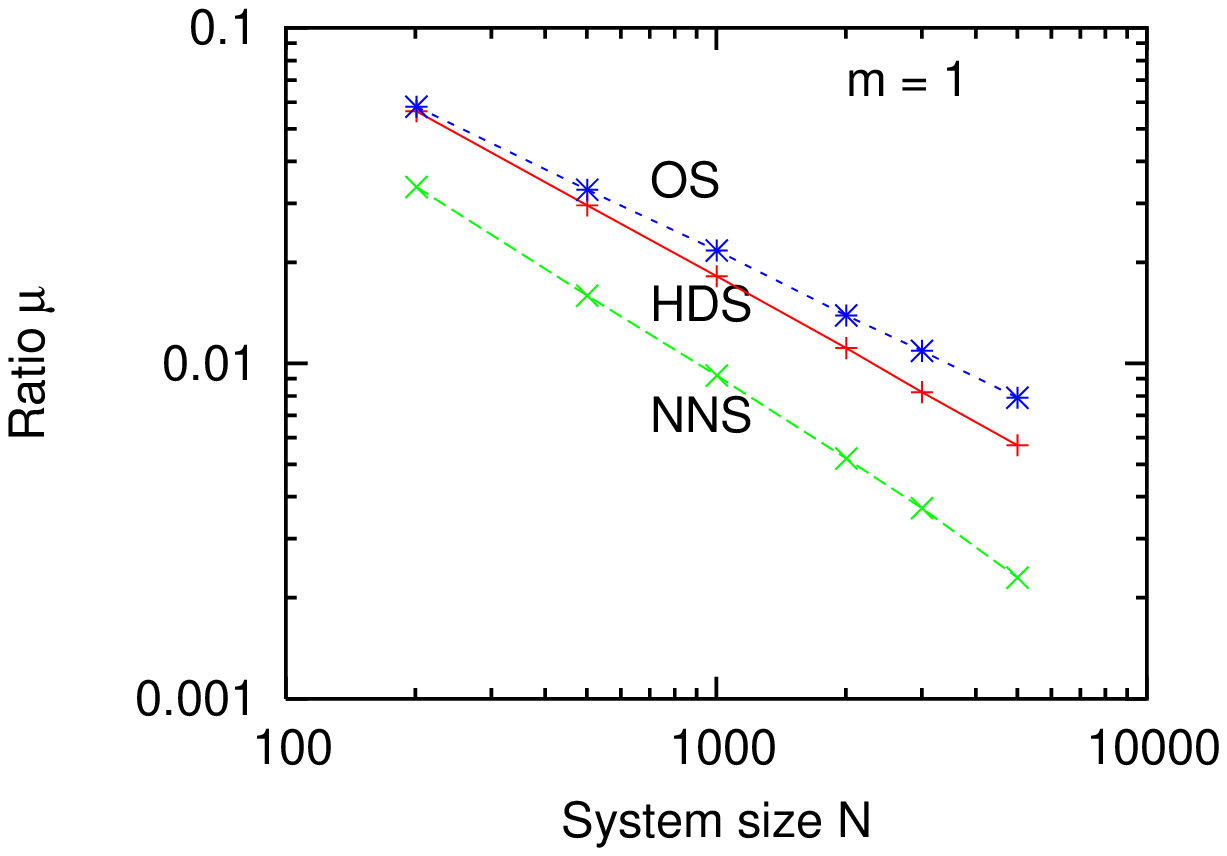}
 \includegraphics[clip,width= 4cm, angle=270]{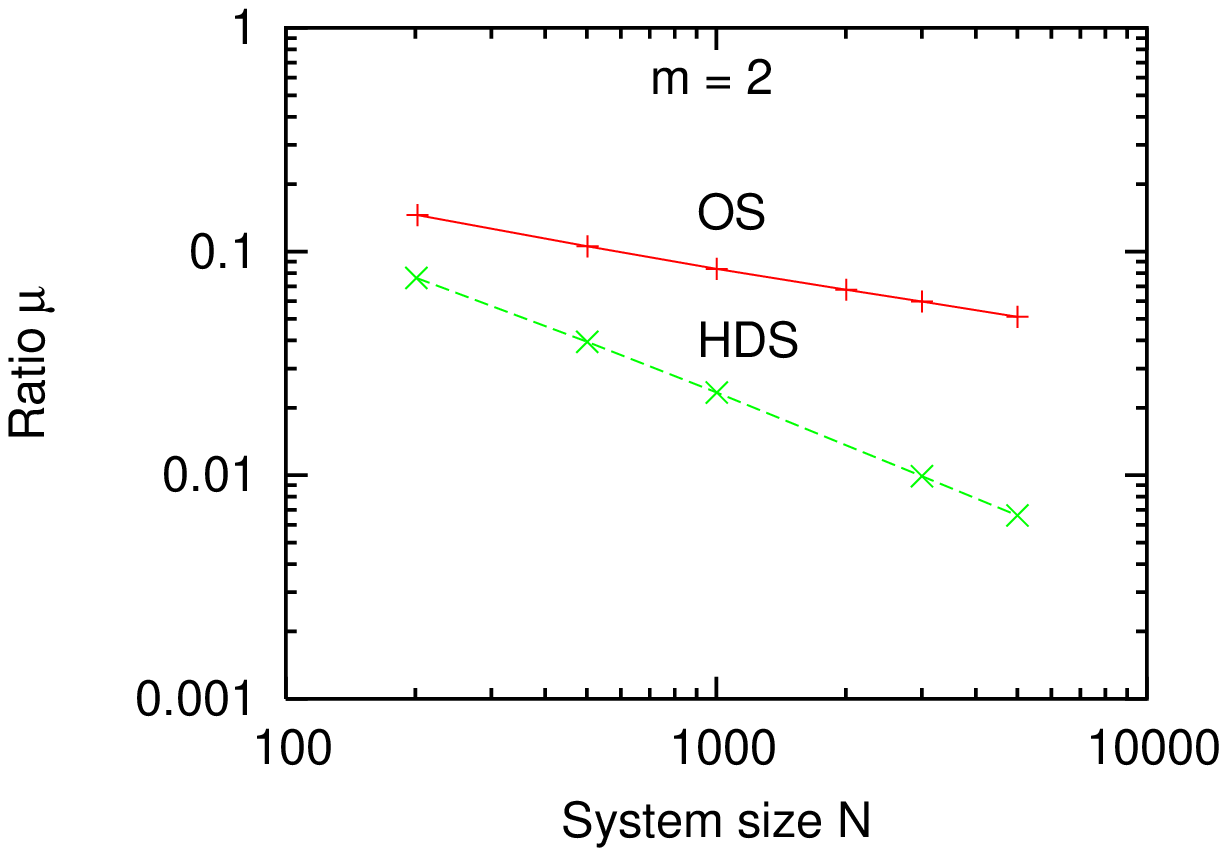}

\caption{The ratio $\mu = \rho/s_d$ vs $N$ plots  at 
$\alpha = 1.0$ for $m=1$ and  $m=2$.}
\end{figure}

 We show typical  plots of $\rho$ and $s_d$ against $N$ (Figs 2,3) for a 
fixed value of 
$\alpha = 1.0$.
$\rho$ clearly shows a power law decay  with $N$. $s_d$  apparently 
has a power law increase, with a very small exponent ($\sim 0.01$). 
However here one 
expects $s_d \sim \ln(N)$ as the network has a tree structure for $m=1$. 
Indeed, 
we find that the exponent tends to decrease at larger $N$, consistent with this.

From Fig. 4, we find that  $\mu$ shows a power law 
variation with the system size $N$, 
\begin{equation}
 \mu \sim N^{-\delta},
\end{equation} 
where the value of $\delta$ varies with  $\alpha$. 
%

\begin{figure}
\noindent \includegraphics[clip,width= 5cm, angle=270]{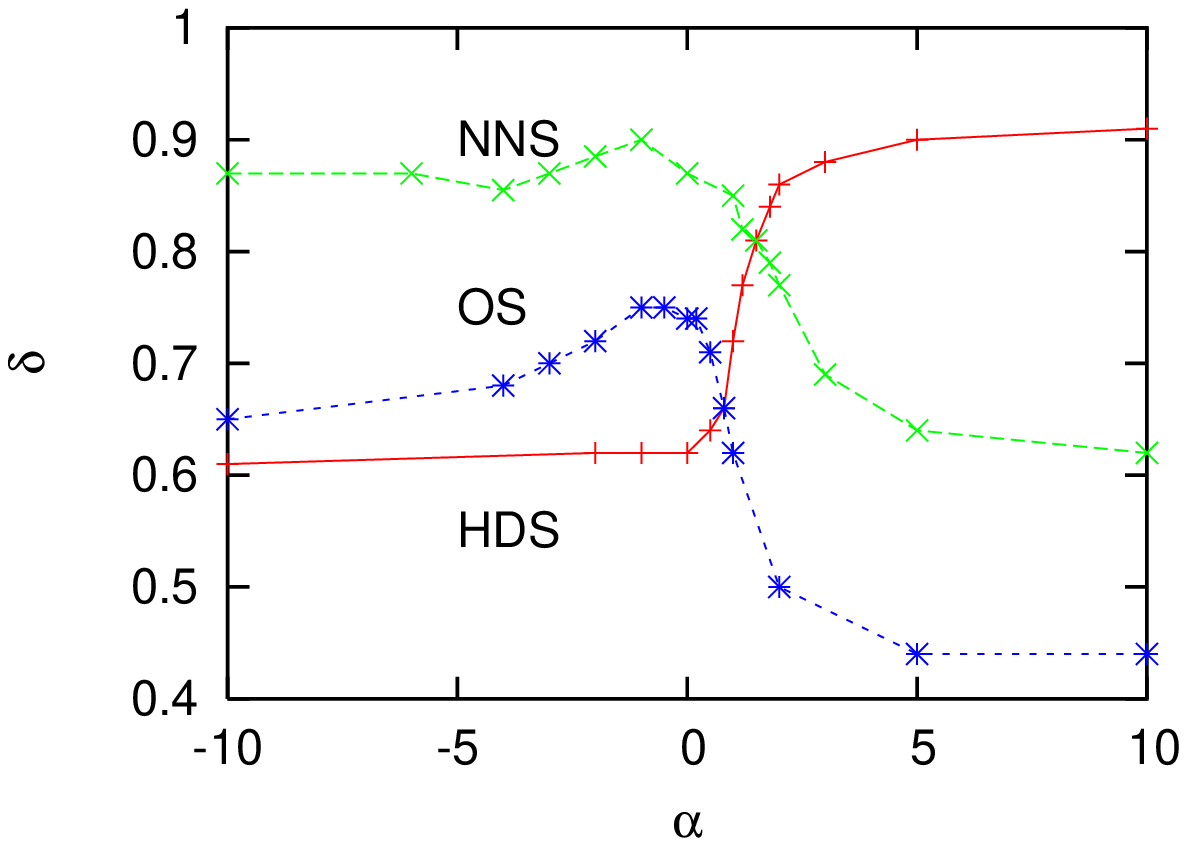}
\caption\protect{\label{fig:transition}Comparison of the different search strategies showing the variation of 
the exponent $\delta$ with $\alpha$. All the three search strategies show 
transitional behaviour close to the static phase transition point $\alpha \simeq 0.5$. $m=1$ here.}
\end{figure}

We have computed $\delta$ for different values of $\alpha$ (from $\alpha = -10.0$ to $\alpha = +10.0$) and shown its 
variation against $\alpha$ for all the three strategies in Fig. 5. 
A smaller value of $\delta$ indicates a more successful  strategy.
For values of $\alpha \leq 0.5$, i.e., when
 the network  is still scale free, the strategy purely  dependent on 
degree works  better compared to   those dependent on  distance. 
For $\alpha >> 0$ on the other hand, the distance dependent 
searches perform better than the 
purely degree dependent search.
In fact, the OS appears to be the best strategy immediately beyond the
static critical point while the NNS works better than the HDS only
when $\alpha  > 1.5$.
 For $\alpha < \alpha_c$ the network is scale free and there are
 several high degree nodes present so that the HDS strategy wins over 
the other two. However this strategy becomes inefficient
 beyond $\alpha > \alpha_c$, when the system is no
 longer scale free  and high degree nodes are no longer available. 
On the  other hand, in this region, distance-based search strategies 
work more efficiently as the network has nodes linked to 
closer neighbours and both the algorithms, NNS and OS  are greedy algorithms 
as far
as distances are concerned.
Although for $\alpha > \alpha_c$, the OS strategy works best,  the exponent 
$\delta$ is never very close to zero, which means that the 
dynamic small world effect \cite{psen} is absent here. For all the strategies however, 
$0 < \delta < 1$,  consistent with the boundary values 
obtained in \cite{psen}. 
It is observed that for a narrow region of values of   $\alpha > \alpha_c$, HDS is 
still better than NNS 
which indicates  that the relevance of the degree of a node reduces
 gradually, once the network becomes non-scale free.  The fact that the
OS performs best even for very large values of $\alpha$ also suggests that
the degree is never totally irrelevant. 

For both HDS and OS, 
$\delta$ shows a drastic increase/decrease, indicating a sharp transition at 
$\alpha \simeq 0.5$, which is the static phase transition point.
Unlike HDS and OS, there is 
no sharp change in behaviour  in $\delta$ for NNS and it is affected by the 
static phase transition point to a lesser extent compared to the other two 
strategies. 

NNS at $\alpha = 0$ is nothing but a random search. $\delta$ for NNS remains 
almost a constant for $-\infty < \alpha < 0$, showing that it is never better 
than a random search strategy for this region. 
For negative values of $\alpha$, 
nodes at large distance are linked up,
but it does not help a greedy algorithm.
 Incorporating $k$ in the algorithm surely helps as OS is better than NNS here.
In fact $\delta$  shows a variation with $\alpha$ for $\alpha < 0$
 only for the OS strategy. For HDS, the plot of $\delta$ versus $\alpha$ 
is close to a perfect sigmoid, showing 
accountable variation only around $\alpha = \alpha_c$. 

\subsection{ Case II, m = 2}

As long as $m=1$, the network cannot have any loop and the path from
one node to another is unique. To introduce loops to the lowest order, we have next
considered 
 searches 
 on    networks generated using (\ref{attach}) once again  
where each incoming node can get two links ($m=2$).
It is expected that the static phase transition point remains same
for $m=2$.
  
With loops,  the success rate should be higher but the search 
lengths may increase several times. 
In the last subsection, we found that the HDS and OS are the
more effective strategies and we have used only these two in the present 
study.
The variation of $\rho$, $s_d$ and $\mu$ with $N$ for $m=2$ 
have been plotted in 
figures 2, 3 and 4 along with the $m=1$ plots. All these show power law variations
with $N$. As expected, we find a slower decay of $\rho$ compared to that
for $m=1$ while $s_d$ increases clearly with a power law compared to the 
logarithmic increase obtained for $m=1$. 
The results show that   
$\delta$ for $\alpha=0.0$ and $\alpha = -1$ are very close for 
OS and HDS; in fact for both these values of $\alpha$, $\delta \simeq 0.7$
for the two strategies. This is greater than that of the 
$m=1$ case. A higher value of $\delta$ indicates a deterioration in 
performance,   the rapid increase in the path lengths being the reason behind
this deterioration.

For $\alpha=1$ on the other hand, when the network is no longer a scale-free 
network, $\delta$ values are drastically different for the two strategies. 
For OS, it is much smaller, $ \sim  0.3$, while for HDS, it is 
around $0.8$. The reason behind this is,  with the OS strategy, the path lengths scale 
with a much smaller exponent in the non scale-free region (Fig. 3).  
   
\section {Distributions}

In the last section, we have compared some search strategies on a
growing Euclidean network by computing the
quantities like success rates and search lengths, where the  mean value of these
quantities have been used to obtain the scaling behaviour.
We have also computed the distributions of these quantities to see whether the
presence of the phase transition has any effect  on these.
The results for three values of $\alpha$, $\alpha< 0$, $\alpha =0$ and 
$\alpha > \alpha_c$ are reported here. 

\begin{figure}
\includegraphics[clip,width= 4cm, angle=270]{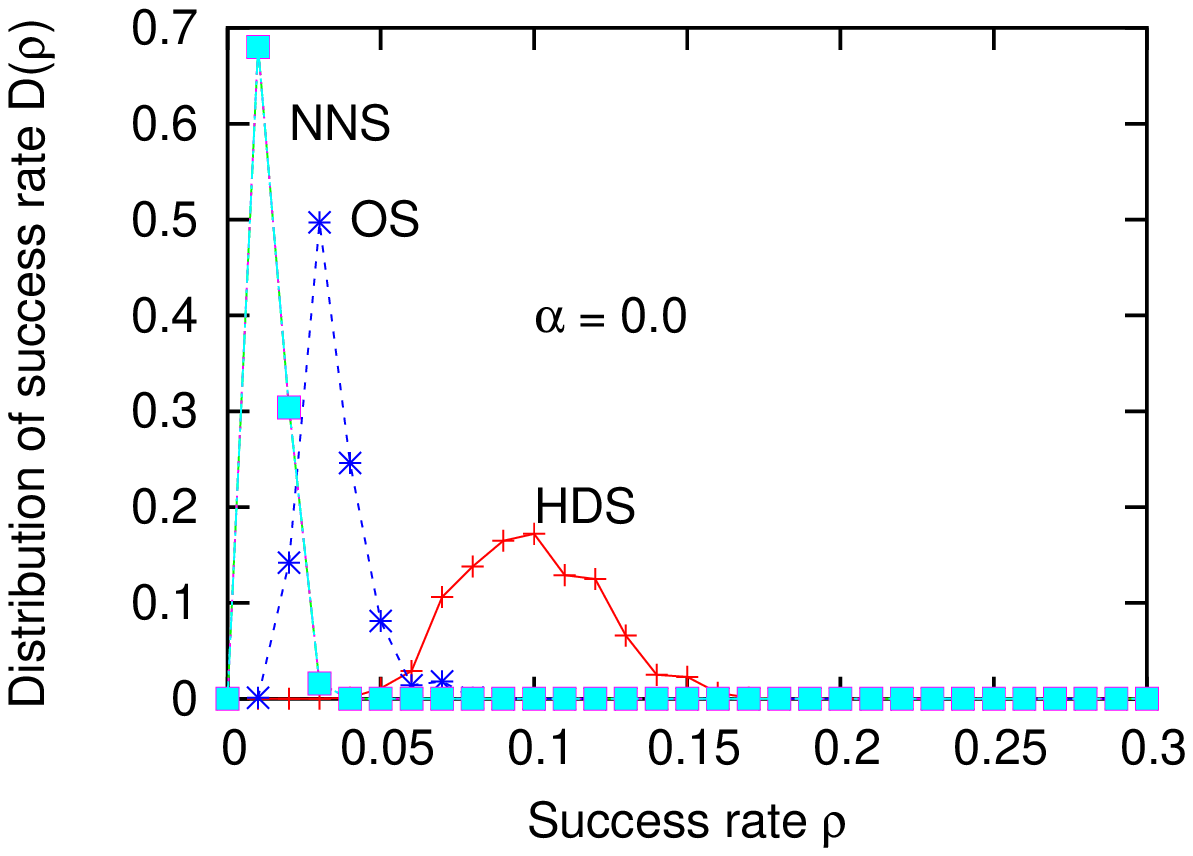}
\includegraphics[clip,width= 4cm, angle=270]{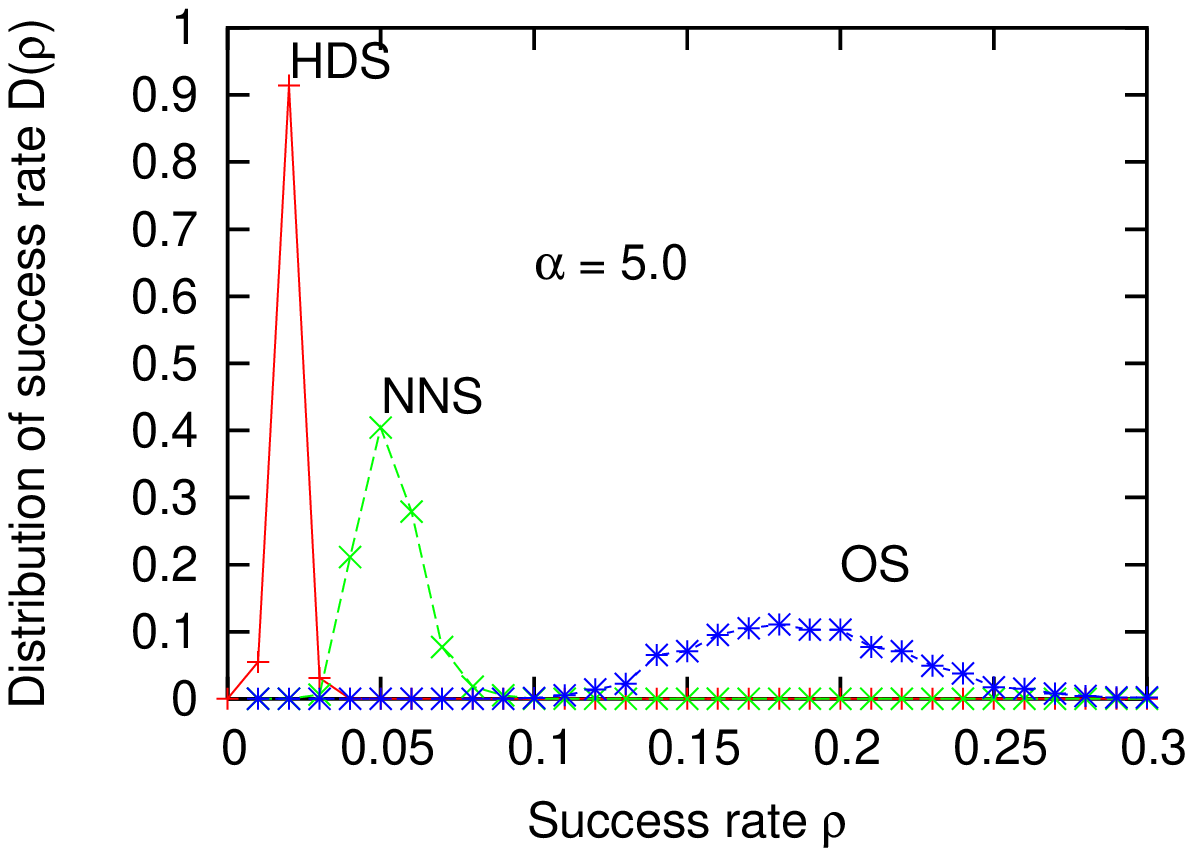}
\includegraphics[clip,width= 4cm, angle=270]{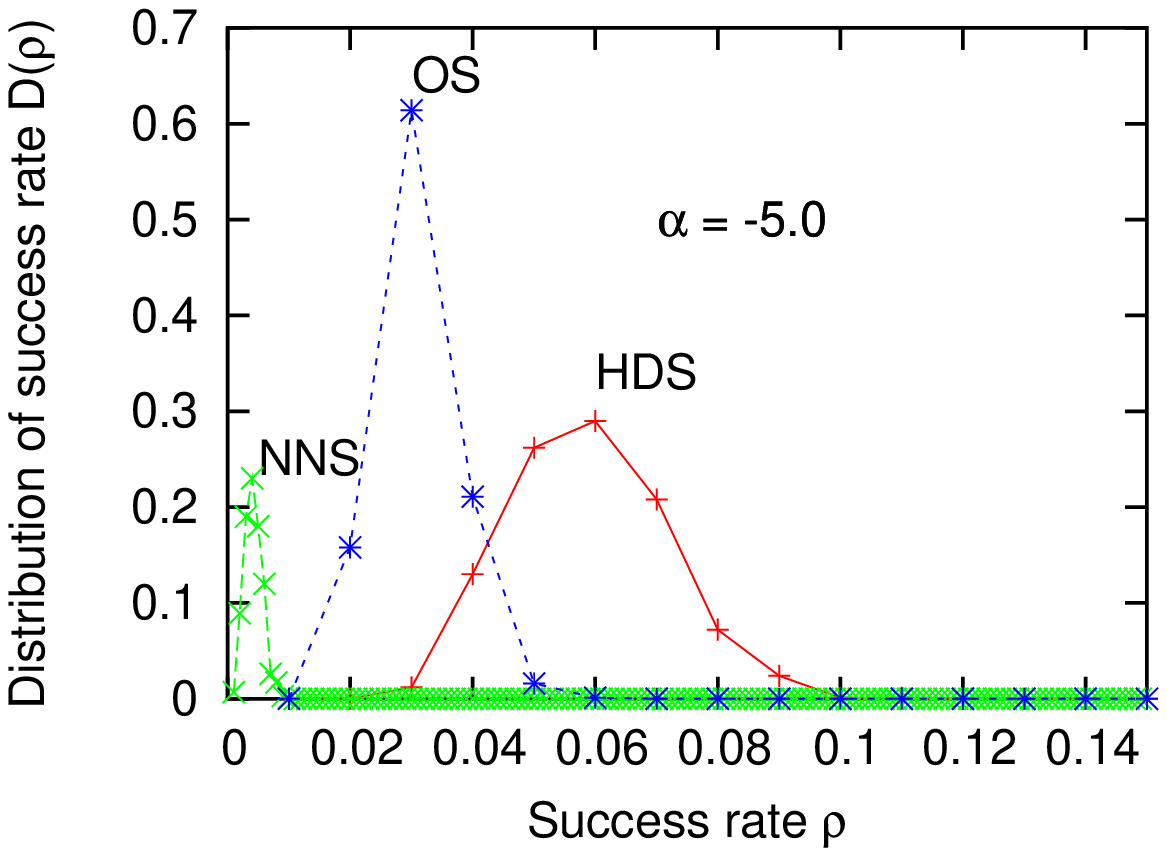}
\caption{Plots of the distributions of $\rho$ for the three strategies for
$\alpha = 0.0$, $5.0$ and $-5.0$ are shown for $m=1$. The system size used is $N = 2000$.}
\end{figure}

The distributions of the success rate $\rho$ for the three strategies with
$m=1$ show the following general features (Fig. 6)\\
1. All of them have a well defined peak.\\
2. They are symmetric.\\
3. There is no long tail.\\
4. Distributions are skewed when the mean value 
is small, in fact very few points with non-zero value appear here.
However, when the mean value is larger, there is a sufficient broadening, no matter
which strategy is being used.

Overall, we do not find any indication that the static phase transition point
has a significant influence on the form of the distributions.
Since the data points are few, we do not attempt a fitting but in all probability these
distributions are gaussian or nearly gaussian.

The distribution of the path length $s_d$, on the other hand, is definitely not 
symmetric (Fig. 7) for any of the strategies at any value of $\alpha$. For $m=1$, it has a broad peak. None of the distributions have a long tail.
For $m=2$, when the success rate becomes much higher, we find the peaks shifted towards lesser
values of $s_d$ (Fig. 8), consistent with the observation of \cite{geog}. The presence of a peak
at smaller values of $s_d$ for both $m=1$ and $m=2$ also show that shorter paths are more probable \cite{dodds}.  

Again, for $m=1$, the number of data  points are few 
and larger fluctuations exist making 
it difficult to fit the data to any  familiar functional form. For $m=2$, 
there is  a 
larger number of points and one can immediately see that the optimised 
 search strategy has a
clear-cut exponential decay when $\alpha > \alpha_c$ whereas for $\alpha < \alpha_c$, it has a slower
than exponential decay. 
However, no such change in behaviour is observed for the HDS strategy, it is  slower than exponential
in each case.

\begin{figure}
\includegraphics[clip,width= 4cm, angle=270]{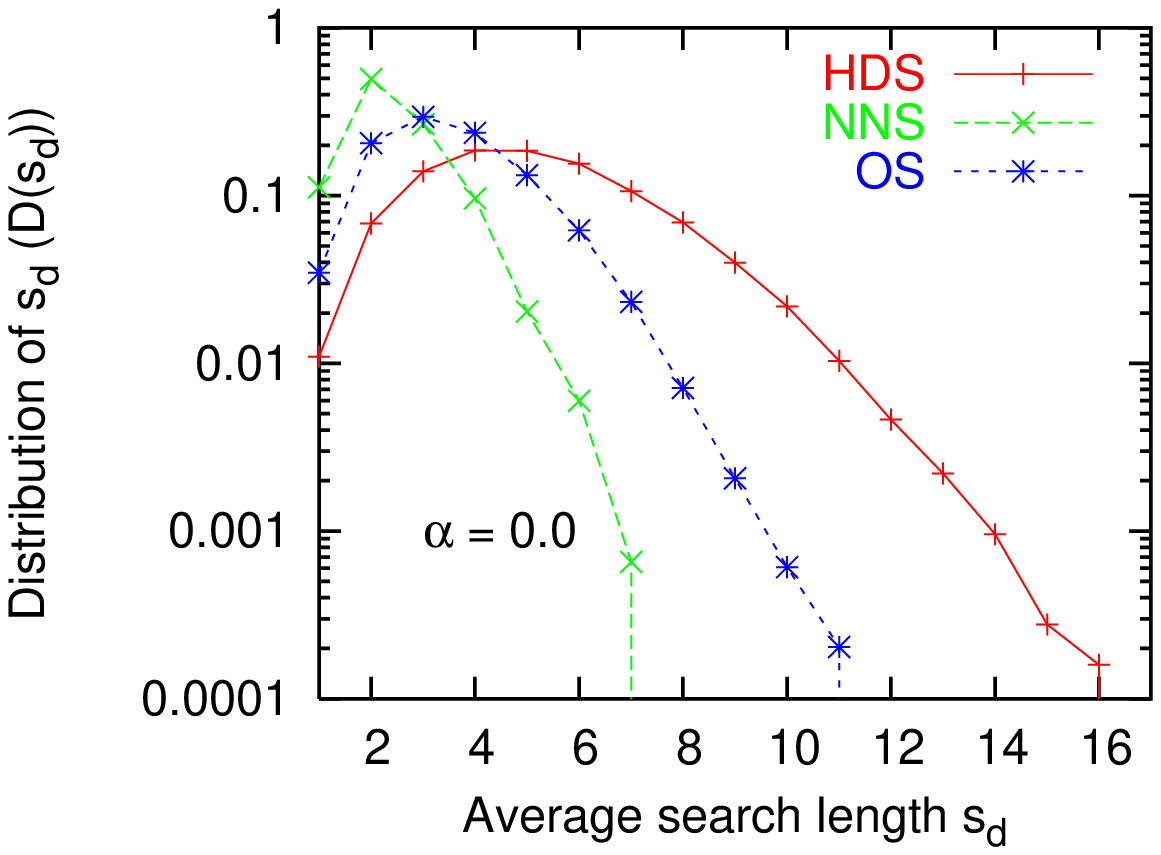}
\includegraphics[clip,width= 4cm, angle=270]{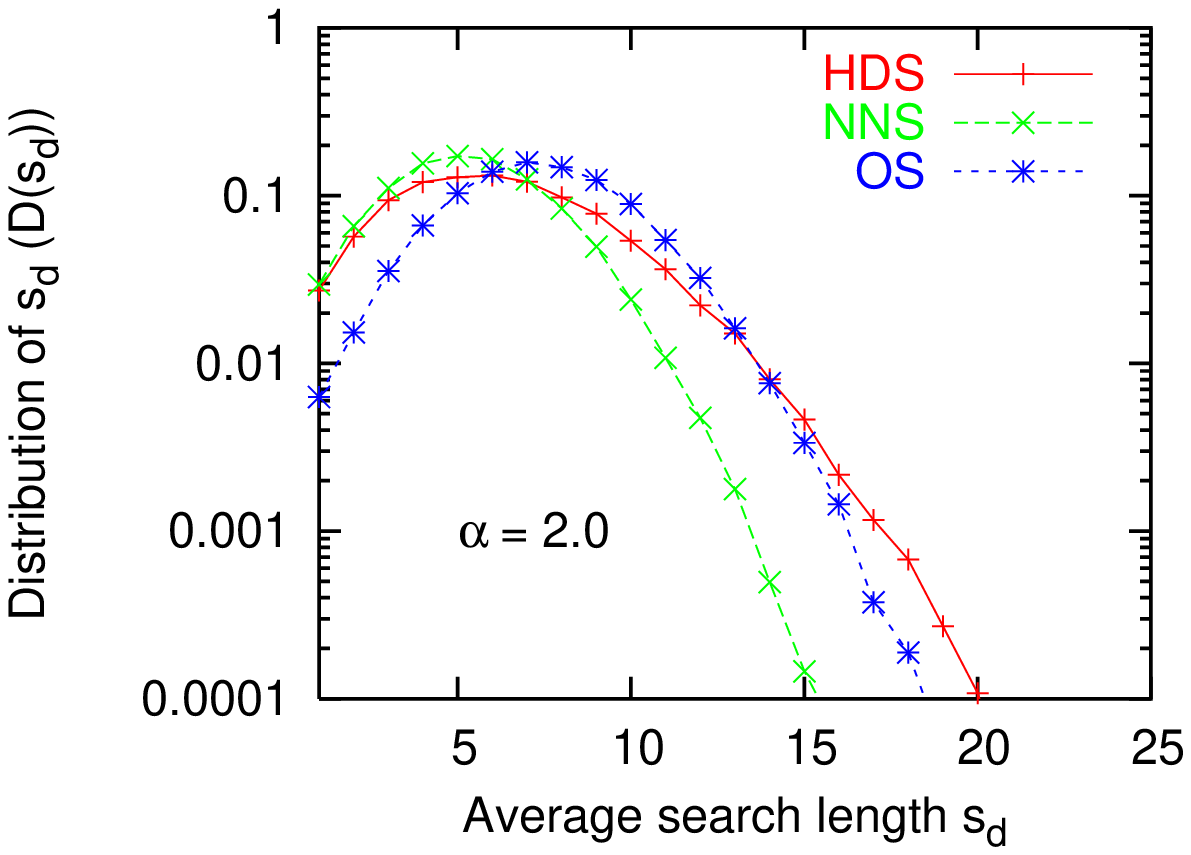}
\includegraphics[clip,width= 4cm, angle=270]{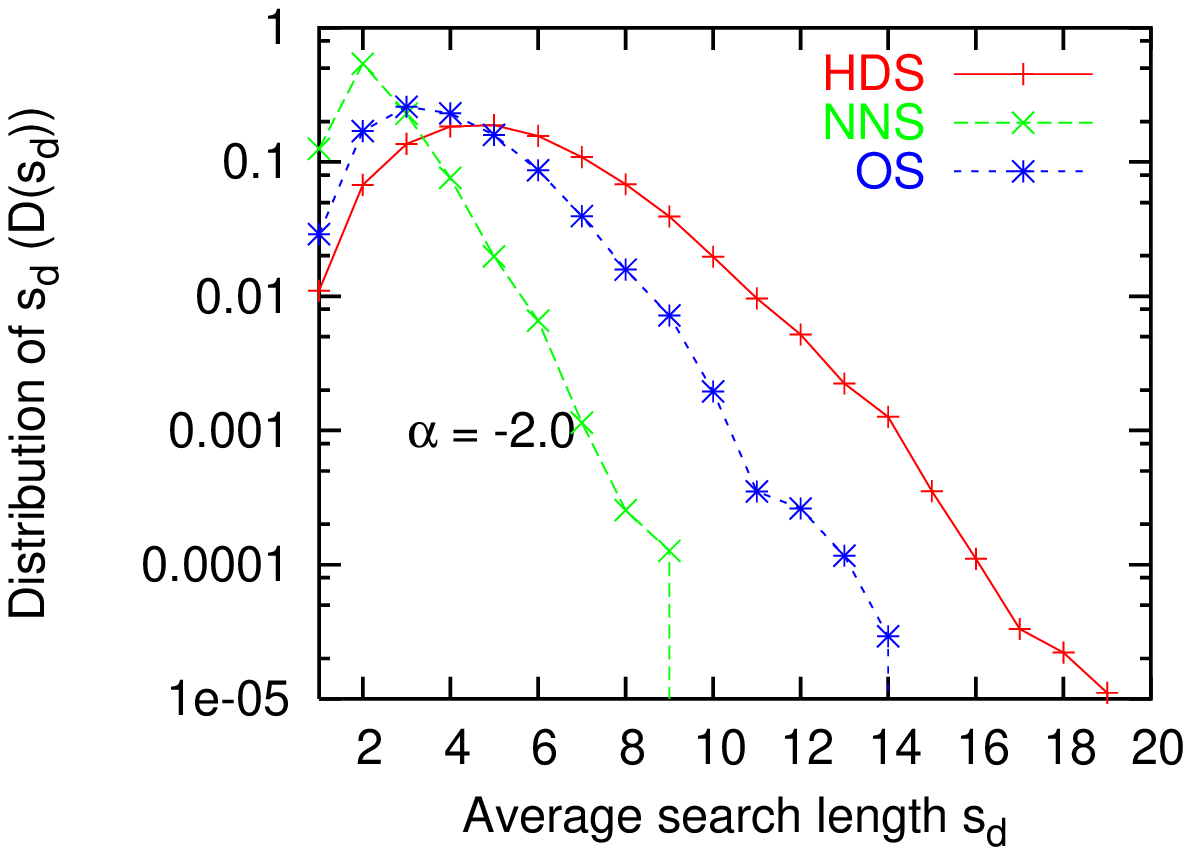}
\caption{The distributions of $s_{d}$ for the three strategies for
$\alpha = 0.0$, $2.0$ and $-2.0$ are shown  in a log-linear scale. The system size used is $N
= 2000$. $m=1$ here.}
\end{figure}


\begin{figure}
\includegraphics[clip,width= 4cm, angle=270]{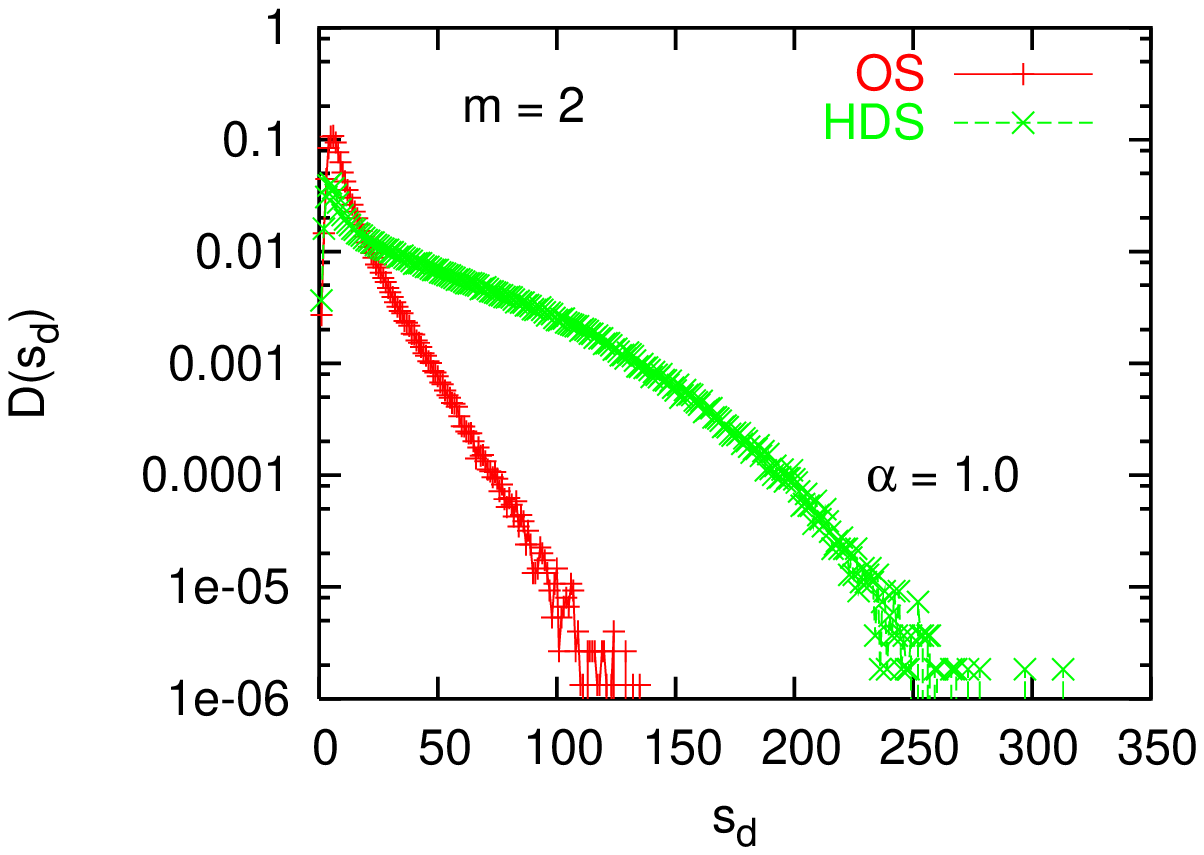}
\includegraphics[clip,width= 4cm, angle=270]{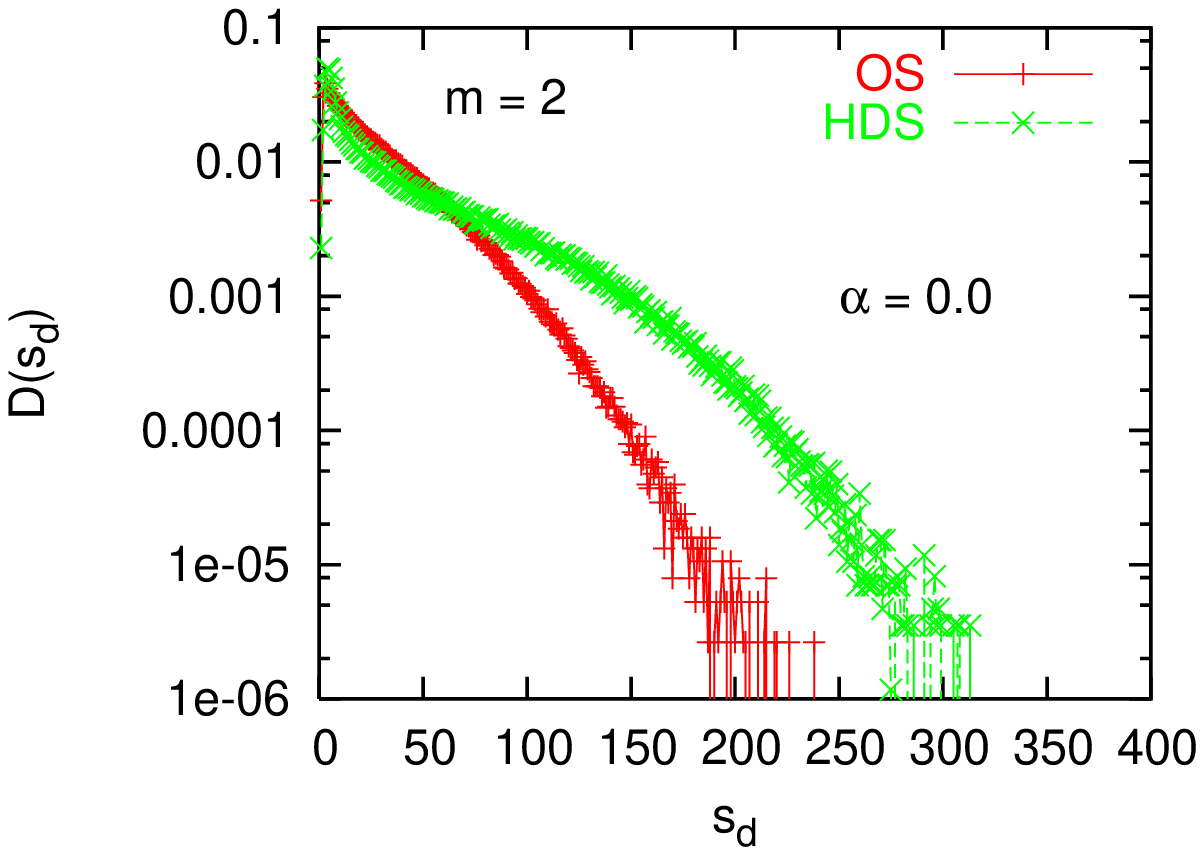}
\includegraphics[clip,width= 4cm, angle=270]{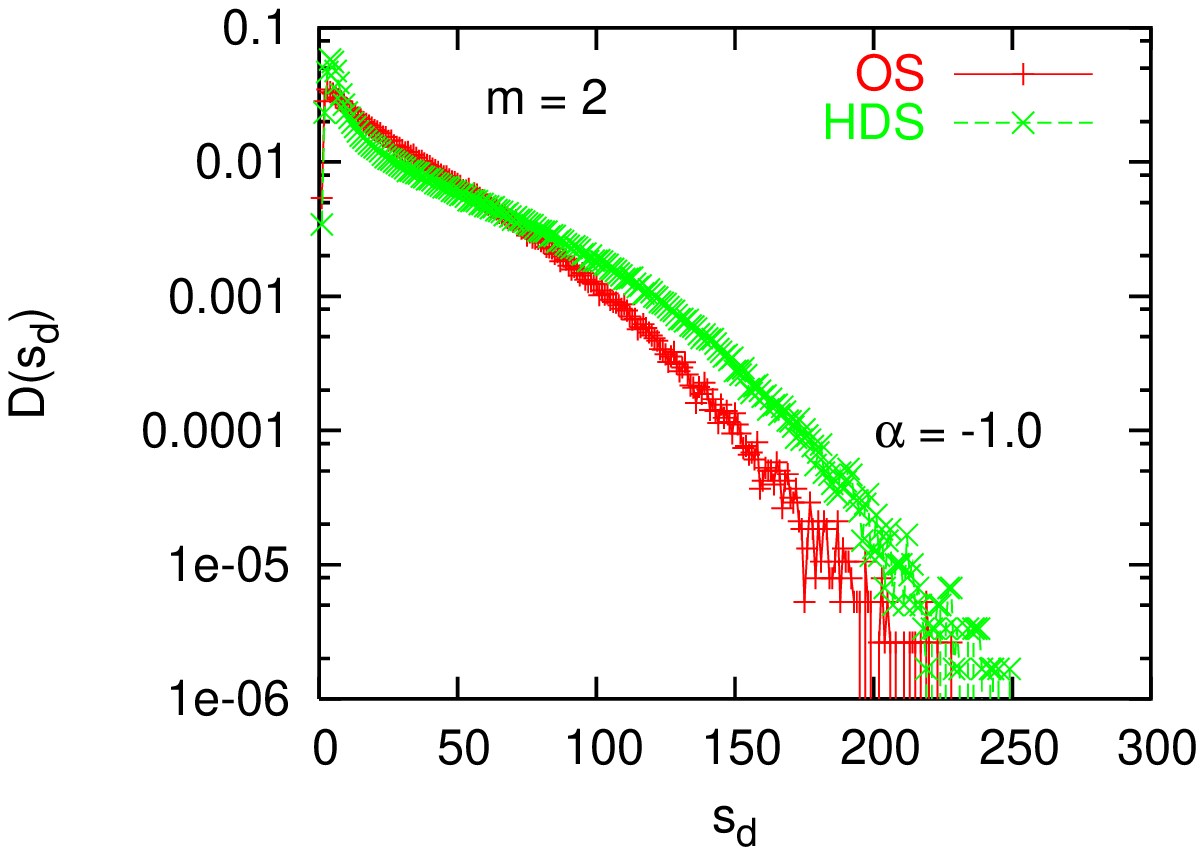}
\caption{The distributions of average search lengths for OS and HDS strategy at 
$\alpha = 1, 0, -1$ when  $m=2$ are shown. $N=2000$ here.}
\end{figure}


\section{Summary and Conclusions}

In this work, we have applied different search strategies  to a network which undergoes  a phase
transition from a scale-free to a non scale-free phase. One of our aims was
to investigate whether such a phase transition significantly affects the 
search or not as in purely Euclidean networks, such an indication is there
\cite{klein,zhu}.
We find that the search strategies indeed show a significant change in behaviour at or near the 
phase transition to different extents. It appears that the degree based searches are
more sensitive to the phase transition. 

The searching scheme used here allows termination of messages as the restriction that
the message can be passed only once by any messenger has been  imposed. 
In reality, of course, several other reasons may exist for a termination \cite{dodds}. The analysis of the results 
therfore has been made   based on an approach recently 
suggested by one of us \cite{psen}, in which both search paths and success rates are taken into consideration.

Searching phenomena is vastly studied in social networks and the present
study also uses ideas relevant to Milgram-like searches. Most social networks being non scale-free,
our results for the network in its non scale-free phase is important in the context of 
social searches.
Here the best performance is shown by the optimised search  (OS) strategy, in which a 
node sends the signal to its neighbour having the largest value of  $k/l$
($k$ is its degree and  $l$ the distance from the target, see sec II).
We have used here three strategies and for none of them  we observe 
a dynamical small world effect, i.e., $\delta$ is never very close to zero. However,
our list of strategies is by no means exhaustive.  
The OS  scheme can be generalised by making the  message passing 
rule  that a  node sends 
the signal to a neighbour with the largest value of 
$k^a/l^b$, introducing tunable parameters $a$ and $b$. 
In this paper, we have only considered the limiting cases 
$a \to \infty$ (HDS), $b \to \infty$ (NNS) and $a=1,b=1$ (OS).
It may be an interesting future study to find out whether in the $a-b$ plane,
one obtains regions of dynamic small world effect.
 
A more detailed study for $m=2$ (or more) can also  be done for which we have presented results
at some specific values of $\alpha$  only. 

We have also estimated the distributions for the success rate and path lengths. 
The static phase transition seems to seriously affect only the distribution for the path lengths 
for the optimised strategy when $m=2$.  

Acknowledgement: Financial support from CSIR grant no. 3(1029)/05-EMR-II
 (PS) and F.no.9/28(609)/2003-EMR-I (KBH)  is  acknowledged.
Computational facility has been partially provided by DST FIST project.

\end{document}